\documentclass[aps,prb,superscriptaddress,showpacs]{revtex4-1}

\usepackage[utf8]{inputenc}					
\usepackage{lmodern}						

\usepackage[dvips]{graphicx}
\usepackage[namelimits]{amsmath}
\usepackage{amsfonts,amssymb,amsthm}

\usepackage[english]{babel}					

\usepackage{mathtools}						
\mathtoolsset{showonlyrefs,showmanualtags}	

\newcommand{\Ai}{{\rm Ai}}
\newcommand{\Bi}{{\rm Bi}}
\providecommand{\abs}[1]{\lvert#1\rvert}

\renewcommand{\figurename}{Fig.}

\begin{document}

\title{Scattering and delay time for 1D asymmetric potentials: the step-linear and the step-exponential cases}
\date{\today}

\author{L. Rizzi}
\affiliation{CNRS, CMAP \'Ecole Polytechnique, and \'Equipe INRIA GECO Saclay \^Ile-de-France, Paris, France}
\author{O. F. Piattella}
\affiliation{Departamento de F\'isica, Universidade Federal do Esp\'irito Santo, avenida Ferrari 514, 29075-910 Vit\'oria, Esp\'irito Santo, Brazil}
\author{S. L. Cacciatori}
\affiliation{Dipartimento di Scienza e Alta Tecnologia, Universit\`a dell'Insubria, Via Valleggio 11, 22100 Como, Italy}
\affiliation{INFN, sezione di Milano, Via Celoria 16, 20133 Milano, Italy}
\author{V. Gorini}
\affiliation{Dipartimento di Scienza e Alta Tecnologia, Universit\`a dell'Insubria, Via Valleggio 11, 22100 Como, Italy}

\begin{abstract}
We analyze the quantum-mechanical behavior of a system described by a one-dimensional asymmetric potential constituted by a step plus (i) a linear barrier or (ii) an exponential barrier. We solve the energy eigenvalue equation by means of the integral representation method, classifying the independent solutions as equivalence classes of homotopic paths in the complex plane.

We discuss the structure of the bound states as function of the height $U_0$ of the step and we study the propagation of a sharp-peaked wave packet reflected by the barrier. For both the linear and the exponential barrier we provide an explicit formula for the delay time $\tau(E)$ as a function of the peak energy $E$. We display the resonant behavior of $\tau(E)$ at energies close to $U_0$. By analyzing the asymptotic behavior for large energies of the eigenfunctions of the continuous spectrum we also show that, as expected, $\tau(E)$ approaches the classical value for $E\to\infty$, thus diverging for the step-linear case and vanishing for the step-exponential one.

%
%
\end{abstract}

\maketitle

\section{Introduction}

In \cite{RPCG2010}, we have analyzed the quantum-mechanical behavior of a system described by a one-dimensional asymmetric potential formed by a step plus a harmonic barrier (the ``step-harmonic'' potential), by using the integral representation method \cite{Hochstadt1976}. We investigated the behavior of the discrete energy levels (as a function of the height of the step) and of the delay time $\tau$ of a wave packet coming from infinity and bouncing back on the harmonic barrier, as a function of the packet's peak energy and of the height $U_0$ of the step.

Among the convex or concave locally bounded symmetric and confining potentials the harmonic oscillator is a threshold, in that it gives rise to classical isochronous oscillations and evenly spaced quantum energy levels \cite{Carinena:2007zz, Asorey:2007gb}.

In our quantum mechanical step variant of the problem we recover both these features in the limit in which $U_0 \to \infty$, and the potential reduces to the half-space harmonic oscillator. Then, it is conceivable that the harmonic one is the only confining barrier which displays a constant nonvanishing delay $\tau$ in the limit of high energies. For steeper barriers we expect $\tau$ to vanish at high energies, while for milder ones we expect the delay to become infinite in this limit, in accordance with the corresponding classical situations. Similarly, we expect that, as $U_0 \to \infty$, the spacing between two neighboring discrete levels tends to infinity in the former case and to zero in the latter.

Here we analyze the ``step-linear'' and the ``step-exponential'' potentials. Both these problems can be solved exactly using the integral representation method, which is interesting \emph{per se}, as it can be applied to more general problems.

\section{The Step-Linear potential}\label{sec:The step-linear potential}

Let $M, U_0$ be positive parameters, and consider the ``step-linear'' potential
\begin{equation}\label{Steplinpot}
 U(x) = \begin{cases}
 -Mx & x \leq 0,\\
 U_0 & x > 0,
\end{cases}
\end{equation}
If $E$ denotes the energy of the particle and $m$ its mass, the time-independent Schr\"odinger equation is
\begin{equation}\label{Schreq}
 -\dfrac{\hbar^2}{2m}\frac{d^2u(x)}{dx^2} + U(x)u(x) = Eu(x) .
\end{equation}
This, for $x < 0$, can be rewritten as
\begin{equation}\label{Schreq2}
\frac{d^2u(x)}{dx^2} + \left(\dfrac{2mM}{\hbar^2}x + \dfrac{2mE}{\hbar^2}\right)u(x) = 0 .
\end{equation}
It is convenient to define
\begin{equation}\label{alphabeta}
\alpha := \left(\dfrac{2mM}{\hbar^2}\right)^{1/3} , \qquad \beta := \dfrac{\alpha E}{M} , \qquad y := \alpha x ,
\end{equation}
which allows us to recast  \eqref{Schreq2} as follows:
\begin{equation}\label{Schreq3}
\frac{d^2u(y)}{d y^2} + \left(y + \beta\right)u(y) = 0 ,
\end{equation}
which is the Airy equation (see \cite{AS1972, Airybook}). The general solution of  \eqref{Schreq3} is
\begin{equation}\label{eigenfunction}
u(y) = C \Ai(-y - \beta) + D \Bi(- y - \beta) ,
\end{equation}
where $C$ and $D$ are arbitrary integration constants, and the two linearly independent solutions $\Ai$ and $\Bi$ are expressed in Section~\ref{App:Airyfun} in terms of the integral representation method.

Since $\Bi(x) \to + \infty$ for $x\to +\infty$, in order for  \eqref{eigenfunction} to be an eigenfunction, we must set $D = 0$. Hence, we obtain
\begin{equation}\label{xl0sol}
u(x) = C\Ai(-\alpha x - \beta) .
\end{equation}
For $x > 0$, the solution of  \eqref{Schreq} has the following from:
\begin{equation}\label{xg0sol}
 u(x) = \begin{cases}
 Ae^{ikx} + Be^{-ikx} &  E > U_0,\\
 Fe^{-k x} & E < U_0,
\end{cases}
\end{equation}
where $A$, $B$ and $F$ are arbitrary integration constants and 
\begin{equation}\label{kappa}
\hbar k := \sqrt{2m\lvert E - U_0\rvert} .
\end{equation}

\subsection{The case $E < U_0$: bound states and level spacing}
\begin{figure}
\centering
\includegraphics[width=0.45\columnwidth]{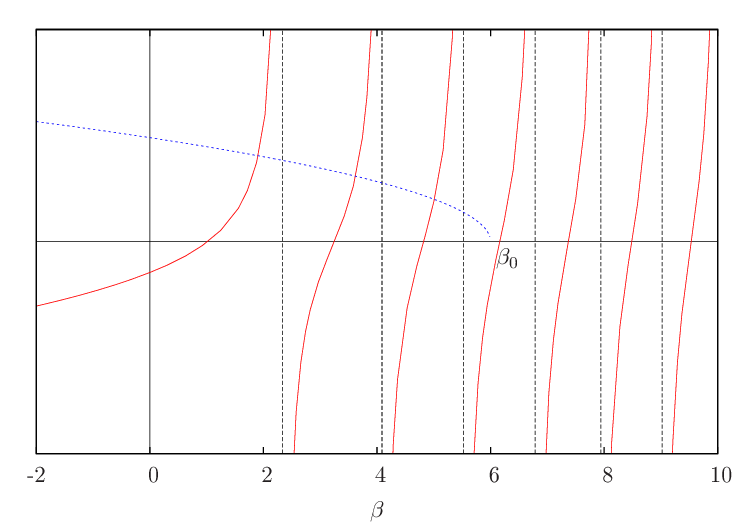}
\caption{Graphical solutions of \protect\eqref{energylevelcond} for the energy levels. The red solid line represents the graph of the function $\Ai'(-\beta)/\Ai(-\beta)$ while the blue dash-dotted line represents the graph of $\sqrt{\beta_0 - \beta}$. In this plot, we have chosen $\beta_0 = 6$.}\label{Fig1}
\end{figure}

If $E < U_0$ we obtain
\begin{equation}\label{WlU0sol}
 u(x) = \begin{cases}
 C \Ai(-\alpha x - \beta)  & x \leq 0,\\
 Fe^{-k x} & x > 0.
\end{cases} 
\end{equation}
The requirement of continuity of $u(x)$ and of its first derivative in $x = 0$ is expressed by
\begin{equation}\label{juncondsWlU0}
\begin{cases}
 C \Ai(-\beta) - F =  0,\\
 C \alpha\Ai'(-\beta) - Fk = 0.
\end{cases}
\end{equation}
System \eqref{juncondsWlU0} has a non trivial solution iff
\begin{equation}\label{energylevelcond}
 \dfrac{\Ai'(-\beta)}{\Ai(-\beta)} = \sqrt{\beta_0 - \beta} ,
\end{equation}
where $\beta_0 := \alpha U_0 /M$. The energy levels are determined graphically by the intersections of the curves at the two sides of  \eqref{energylevelcond}. An example is depicted in \figurename~\ref{Fig1} for $\beta_0 = 6$.
In the limit $U_0 \to \infty$ the step of  \eqref{Steplinpot} becomes an infinite barrier. In this case, the energy levels correspond to the zeros $\beta_n$ of $\Ai(-\beta)$, the denominator of  \eqref{energylevelcond}. As expected, these energy levels are the ones of the symmetric confining potential $U(x) = M|x|$ corresponding to the odd eigenfunctions of the latter (Appendix~\ref{App:linearwell}).

To study the level separation for large energies, consider the asymptotics of $\Ai(-\beta)$ for large values of $\beta$ (see \cite{FO99}):
\begin{equation}\label{Aiasympexp}
\Ai(-\beta) = \dfrac{1}{\sqrt{\pi}\beta^{1/4}}
\left[\sin\left(\zeta + \dfrac{\pi}{4}\right)\sum_{k = 0}^{\infty}(-)^kc_{2k}\zeta^{-2k} - \cos\left(\zeta + \dfrac{\pi}{4}\right)\sum_{k = 0}^{\infty}(-)^kc_{2k + 1}\zeta^{-2k - 1}\right] ,
\end{equation}
where $\zeta : = 2\beta^{3/2}/3$ and the coefficients of the series expansions are given by
\begin{equation}\label{coeffexp}
 c_k = \frac{\Gamma\left(3k + \frac{1}{2}\right)}{54^k k! \Gamma\left(k + \frac{1}{2}\right)} .
\end{equation}

The inversion of the asymptotic expansion \eqref{Aiasympexp} allows us to find, for large values of $\beta$, the following approximate solution of the equation $\Ai(-\beta) = 0$ (see \cite{FO99}):
\begin{equation}\label{Texpans}
 \beta_n \sim t_n^{2/3}\left(1 + \frac{5}{48}\frac{1}{t_n^2} - \frac{5}{36}\frac{1}{t_n^4} + \ldots\right) , \qquad t_n = \frac{3}{8}\pi\left(4n - 1\right), \qquad n \to \infty .
\end{equation}
Thus, at the leading order of  \eqref{Texpans} the approximate zeros have the form
\begin{equation}\label{Aizeroes}
\beta_n \simeq \left[\dfrac{3}{8}\pi\left(4n - 1\right)\right]^{2/3} .
\end{equation}
This is an excellent approximation to the zeros of $\Ai(-x)$ (see Table~\ref{tab1}).
\begin{table}[b]
\centering
\begin{tabular}{|c|c|c|c|}
\hline $n$ & $\beta_n$ (exact) & $\beta_n$ (approximate) & Relative Error \\ 
\hline 1 & $2.33811$ & $ 2.32025$ & $0.76 \times 10^{-2}$ \\ 
\hline 2 & $4.08794$ & $ 4.08181$ & $0.15\times 10^{-2}$ \\ 
\hline 3 & $5.52055$ & $ 5.51716$ & $0.62\times 10^{-3}$ \\ \hline
\end{tabular} 
\caption{The first three zeros of the Airy function compared with the corresponding approximate zeros from \protect\eqref{Aizeroes}.}\label{tab1} 
\end{table}
For $n \to \infty$ we get,
\begin{equation}\label{levspacinglin}
\beta_{n + 1} - \beta_n \sim \left(\frac{8}{3n}\right)^{1/3}\pi^{2/3} , \qquad n\to\infty .
\end{equation}
The spacing behavior $n^{-1/3}$ is the threshold between concave and convex potentials.

\subsection{The case $E > U_0$: scattering and delay}

In this case, the (improper) eigenfunctions have the form
\begin{equation}\label{WgU0sol}
u(x) = \begin{cases}
C \Ai(-\alpha x - \beta) & x \leq 0,\\
Ae^{i kx} + Be^{-i kx} & x > 0.
\end{cases} 
\end{equation}
The junction conditions in $x = 0$ are:
\begin{equation}\label{juncondsWgU0}
\begin{cases}
 C \Ai(-\beta) =  A + B,\\
 -C \alpha\Ai'(-\beta) = i k\left(A - B\right).
\end{cases}
\end{equation}
Solving for the constants, the normalized (with respect to $k$) improper eigenfunctions are given by:
\begin{equation}\label{improper_eigenfunctions_lin}
u_k(x) = \frac{1}{\sqrt{2\pi}}\begin{cases}\Pi[\beta(k)]\Ai[-\alpha x -\beta(k)] & x\leq 0,\\
e^{-ikx} + e^{ikx + i\delta(k)} & x > 0,
\end{cases}
\end{equation}
where
\begin{equation}\label{coeffuk}
\Pi(\beta) = 2\left[ \Ai(-\beta) + \frac{\alpha}{ik}\Ai^\prime(-\beta)\right]^{-1} , \qquad e^{i\delta(k)} = \frac{ik\Ai(-\beta) - \alpha\Ai^\prime(-\beta)}{ik\Ai(-\beta) + \alpha \Ai^\prime(-\beta)} .
\end{equation} 
As expected, the continuous part of the spectrum ($E > U_0$) is simple. Note that
\begin{equation}\label{phaseshift}
\delta(k) = 2\arctan\left[\frac{\alpha}{k}\frac{\Ai^\prime[-\beta(k)]}{\Ai[-\beta(k)]}\right] .
\end{equation}
From  \eqref{improper_eigenfunctions_lin} a generic wave packet
\begin{equation}
\psi(x,t) = \int_0^{\infty}d k  c(k) u_{k}(x) e^{-\tfrac{i}{\hbar}E(k) t} .
\end{equation}
has the form
\begin{equation}
\psi(x,t) =  \frac{1}{\sqrt{2\pi}} \begin{cases}
             \int_{0}^{\infty}d k  c(k) \Pi[\beta(k)] \Ai[-\alpha x -\beta(k)] e^{-\tfrac{i}{\hbar}E(k) t} & x < 0 ,\\
\int_{0}^{\infty}d k  c(k) \left[e ^{ikx+i\delta(k)} + e ^{-ikx}\right]e^{-\tfrac{i}{\hbar}E(k) t} = \psi_{\text{refl}}+\psi_{\text{in}} & x > 0.
            \end{cases}
\end{equation}
Then, writing $c(k) = \abs{c(k)}e^{i\gamma(k)}$, $\psi_{\text{in}}$ and $\psi_{\text{refl}}$ take the following form:
\begin{align}
\psi_{\text{in}}(x,t) & =   \frac{1}{\sqrt{2\pi}} \int_{0}^{+\infty}d k  \abs{c(k)} e^{-i[kx+\Omega(k) t - \gamma(k)]} , \\
\psi_{\text{refl}}(x,t) & =   \frac{1}{\sqrt{2\pi}} \int_{0}^{+\infty}d k  \abs{c(k)} e^{i[kx-\Omega(k) t + \delta(k)+ \gamma(k)]} ,
\end{align}
where
\begin{equation}\label{OmegakE}
\Omega(k) := \frac{E(k)}{\hbar} = \frac{U_0}{\hbar} + \frac{\hbar k^2}{2m} .
\end{equation}
If $c(k)$ is sufficiently regular and non-vanishing only in a small neighborhood of some $\tilde{k}$, then $\psi_{\text{in}}$ and $\psi_{\text{refl}}$ represent wave packets which move according to the following equations of motion \cite{Prosperi, griffiths}:
\begin{equation}
x_{\text{in}} = -\left.\dfrac{d \Omega}{d k}\right|_{k=\tilde{k}}t +\left.\dfrac{d \gamma}{d k}\right|_{k=\tilde{k}} = - \dfrac{\hbar \tilde{k}}{m} (t-t_0)= -\dfrac{\tilde{p}}{m} (t-t_0)  ,
\end{equation}
for the ``incoming'' wave packet, and
\begin{equation}
x_{\text{refl}} = \left.\dfrac{d \Omega}{d k}\right|_{k=\tilde{k}} t -\left.\dfrac{d \gamma}{d k}\right|_{k=\tilde{k}} - \left.\dfrac{d \delta}{d k}\right|_{k=\tilde{k}}= \dfrac{\tilde{p}}{m} \left[(t-t_0)-\dfrac{m}{\tilde{p}}\left.\dfrac{d \delta}{d k}\right|_{k=\tilde{k}}\right] ,
\end{equation}
for the reflected ``outgoing'' one.

The solution thus built represents a particle of well defined momentum $\tilde{p}=\hbar\tilde{k}$ which approaches the origin from the right, interacts with the linear potential (at $t = t_0$), and is totally reflected. Note that the argument of the complex valued function $c(k)$ determines $t_0$. The phase shift results in a delay $\tau$ in the rebound, caused by the interaction with the confining linear barrier. The delay is calculated with respect to the case of instantaneous reflection, which takes place in presence of an infinite barrier and for which $\delta(k) = \pi$. From \eqref{alphabeta} and \eqref{kappa} it follows that
\begin{equation}
\tau(\tilde{\beta}) = \frac{\alpha\hbar}{M}\left.\frac{d\delta}{d\beta}\right|_{\beta=\tilde{\beta}}  ,
\end{equation}
where $\tilde{\beta} := \beta(\tilde{k})$.
We compute $\tau$ from  \eqref{phaseshift}. Using the Airy equation $\Ai''(-\beta) = -\beta\Ai(-\beta)$, we obtain
\begin{equation}\label{phaseshiftbeta3}
\tau(\beta) = \frac{\alpha\hbar}{M} \dfrac{2}{\sqrt{\beta - \beta_0}\dfrac{\Ai(-\beta)}{\Ai'(-\beta)} + \frac{1}{\sqrt{\beta - \beta_0}}\dfrac{\Ai'(-\beta)}{\Ai(-\beta)}}\left[-\dfrac{1}{2(\beta - \beta_0)} +  \beta\dfrac{\Ai(-\beta)}{\Ai'(-\beta)} + \dfrac{\Ai'(-\beta)}{\Ai(-\beta)}\right] ,
\end{equation}
where we have suppressed the tilde on the packet peak energy $\tilde{\beta}$.

We are interested in the behavior of the interaction time for large values of $\beta$, i.e. for incoming packets with high energy. To this end, we need the asymptotic expansion of $\Ai'(-\beta)$ for $\beta \to +\infty$ (see \cite{AS1972, FO99}):
\begin{equation}\label{Aipasympexp}
\Ai'(-\beta) = -\dfrac{\beta^{1/4}}{\sqrt{\pi}}
\left[\cos\left(\zeta + \dfrac{\pi}{4}\right)\sum_{k = 0}^{\infty}(-)^k\frac{d_{2k}}{\zeta^{2k}} + \sin\left(\zeta + \dfrac{\pi}{4}\right)\sum_{k = 0}^{\infty}(-)^k\frac{d_{2k + 1}}{\zeta^{2k + 1}}\right] ,
\end{equation}
where $\zeta : = 2\beta^{3/2}/3$ and the coefficients $d_k$ are
\begin{equation}\label{doeffexp}
 d_k = -\frac{6k + 1}{6k - 1}c_k ,
\end{equation}
with $c_k$ given in \eqref{coeffexp}. Dividing the two asymptotic expansions of $\Ai'(-\beta)$ and $\Ai(-\beta)$ we obtain to leading order in $\beta$
\begin{equation}
 \dfrac{1}{\sqrt{\beta - \beta_0}} \dfrac{\Ai'(-\beta)}{\Ai(-\beta)} \sim \tan\left(\dfrac{2}{3}\beta^{3/2} - \dfrac{\pi}{4}\right) , \qquad \beta \to \infty .
\end{equation}
Thus,  \eqref{phaseshiftbeta3} becomes
\begin{equation}
\tau(\beta) \sim 2\frac{\alpha\hbar}{M}\sqrt{\beta} , \qquad \beta \to \infty .
\end{equation}
Hence, reintroducing the physical variables, the high-energy behavior of the interaction time is
\begin{equation}\label{delay_linear_asympt}
\tau(E)  \sim \frac{2\sqrt{2 m E}}{M} ,
\end{equation}
which is exactly the time a classical particle arriving from infinity with energy $E$ would spend in the $x < 0$ region.
\begin{figure}[htbp]
\centering
\includegraphics[width=0.33\columnwidth]{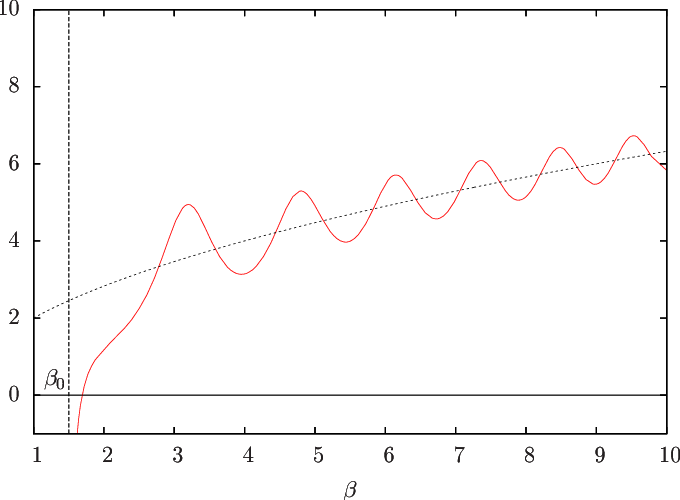}\includegraphics[width=0.33\columnwidth]{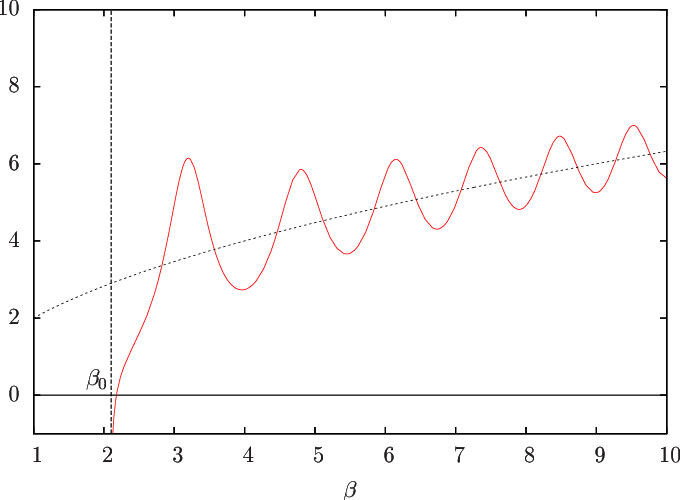}\includegraphics[width=0.33\columnwidth]{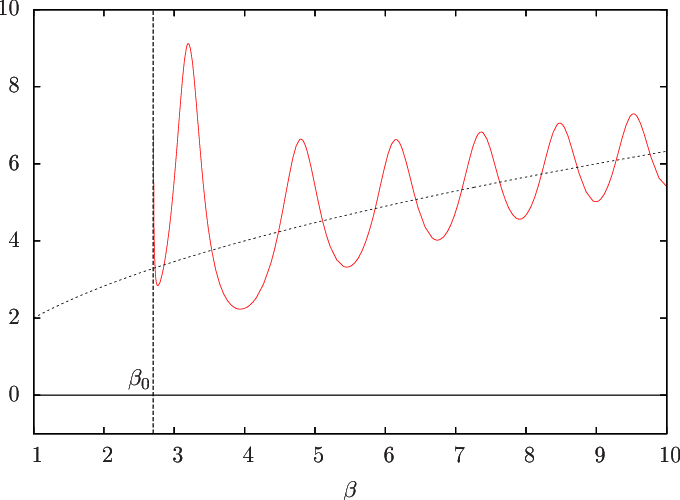}\\
\includegraphics[width=0.33\columnwidth]{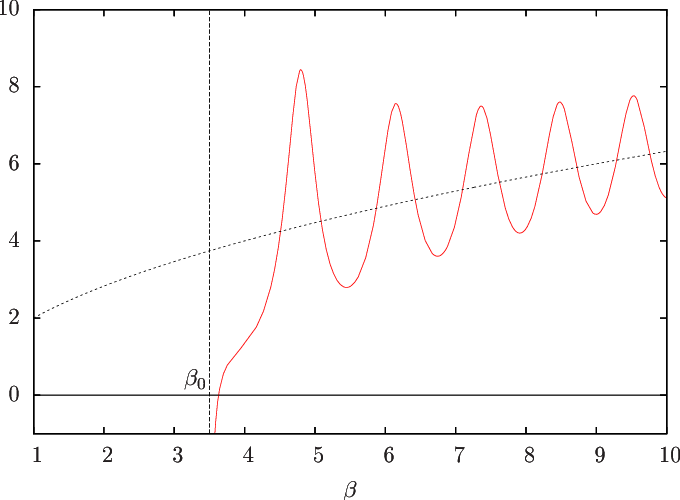}\includegraphics[width=0.33\columnwidth]{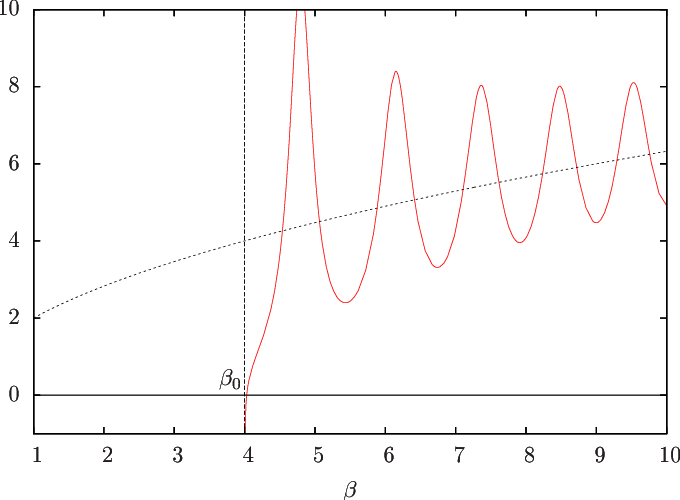}\includegraphics[width=0.33\columnwidth]{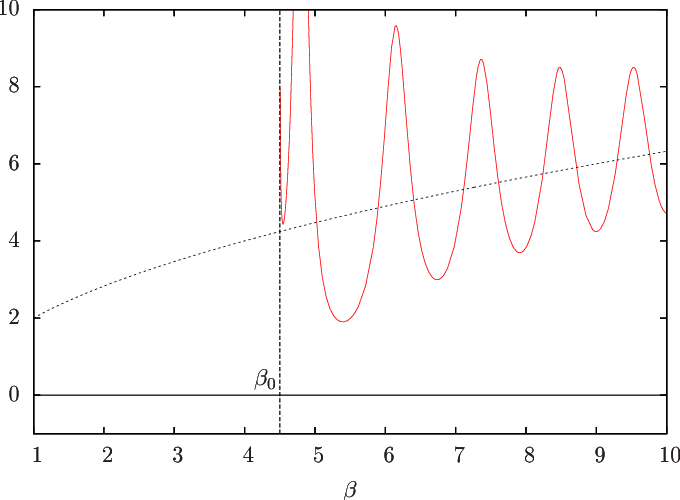}
\caption{Plots of the delay $\tau$ (in units of $\hbar\alpha/M$) versus the energy $\beta$ of the incoming wave packet (red solid lines) for $\beta_0 = 1.5, 2.1, 2.7,3.5, 4.0, 4.5$. The parabolic blue dotted lines represent the classical delay, given in  \protect\eqref{delay_linear_asympt}.}\label{Figrislin}
\end{figure}
In Fig.~\ref{Figrislin} we plot $\tau(\beta)$ for a choice of different values of $\beta_0$. Note the resonances located at the points $\beta \simeq \eta_n$ ($n = 1,2,\dots$), zeros of ${\rm Ai}^\prime(-\beta)$, corresponding to the formation of metastable states at the respective energies $E_n \simeq M\eta_n/\alpha$. The values $M\eta_n/\alpha$ are the energies of the excited states of the confining potential $M|x|$ (see Appendix~\ref{App:linearwell}), corresponding to even eigenfunctions. The resonances have lifetimes which decrease as the corresponding energies increase and move farther away from the threshold energy $U_0$. Conversely, as $U_0$ increases, the lifetime of the resonance closest to the height of the step becomes progressively longer and then infinite when the resonance turns into the next bound state. This behavior is evident in Fig.~\ref{Figrislin}, in which the first three plots correspond to values of $\beta_0$ for which there is only one bound state. In the successive three plots the resonance at $\beta \simeq \eta_1$ has disappeared, having turned into the second bound state. 

Comparing Fig.~\ref{Figrislin} with figure 6 of \cite{RPCG2010} we note that, whereas in the step-harmonic case the graph of $\tau(\beta)$ oscillates with decreasing amplitude about the straight line $\tau = \pi/\omega$ (the half period of the oscillator), in the step-linear case the corresponding graph similarly oscillates about the parabolic line $\tau(E) = 2\sqrt{2mE}/M$, corresponding to the delay of the classical particle. Furthermore, whereas in the step-harmonic case the resonances are evenly spaced, in the step-linear case their spacing decreases with the energy, corresponding to the behavior as a function of the energy of the eigenvalues of the corresponding (symmetric) potentials $U(x) = m\omega^2x^2/2$ and $U(x) = M|x|$.

\section{The Step-Exponential potential}\label{Sec:Step-exponential-potential}

Let $\kappa$, $\sigma$ and $U_0$ be positive parameters, and consider the ``step-exponential'' potential
\begin{equation}\label{Stepexp}
 U(x) = \begin{cases}
 \kappa\left(e^{-x/\sigma} - 1\right) & x \leq 0,\\
 U_0 & x > 0,
\end{cases}
\end{equation}
For $x < 0$, introduce the following dimensionless quantities:
\begin{equation}\label{alphabetastepexp}
\alpha^2: = \frac{8m\kappa\sigma^2}{\hbar^2} , \qquad \beta := \frac{8m(E + \kappa)\sigma^2}{\hbar^2} , \qquad z := \alpha e^{-x/2\sigma} ,
\end{equation}
in terms of which the time-independent Schr\"odinger equation writes as
\begin{equation}\label{Schreq_exp3}
z^2\dfrac{d^2u(z)}{d z^2} + z\dfrac{d u(z)}{d z} + \left(\beta - z^2\right)u(z) = 0 .
\end{equation}
Setting $\nu^2 := -\beta$,  \eqref{Schreq_exp3} can be cast in the form of a modified Bessel equation (see \cite{Watson, AS1972}):
\begin{equation}\label{Schreq_exp4}
z^2\dfrac{d^2u(z)}{d z^2} + z\dfrac{d u(z)}{d z} - \left(\nu^2 + z^2\right)u(z) = 0 ,
\end{equation}
whose general solution is
\begin{equation}\label{gensol}
 u(z) = C K_\nu(z) + D I_\nu(z) ,
\end{equation}
where $C$ and $D$ are arbitrary integration constants and $K_\nu$ and $I_\nu$ are the modified Bessel functions of order $\nu = i\sqrt{\beta}$. 

The function $I_\nu(z)$ diverges exponentially for $z\to+\infty$ \cite{AS1972}. For this reason, in order for $u(x)$ to be a proper (or improper) eigenfunction, we must set $D = 0$. Therefore  \eqref{gensol} reduces to
\begin{equation}\label{xl0sol_exp}
 u(x) = C K_{i\sqrt{\beta}}\left(\alpha e^{-x/2\sigma}\right) .
\end{equation}
Also the solutions of  \eqref{Schreq_exp4} can be studied with the integral representation method (see Section~\ref{App:modbesselfun}).

\subsection{The case $E < U_0$: bound states}

\begin{figure}
 \begin{center}
 \includegraphics[scale=0.7]{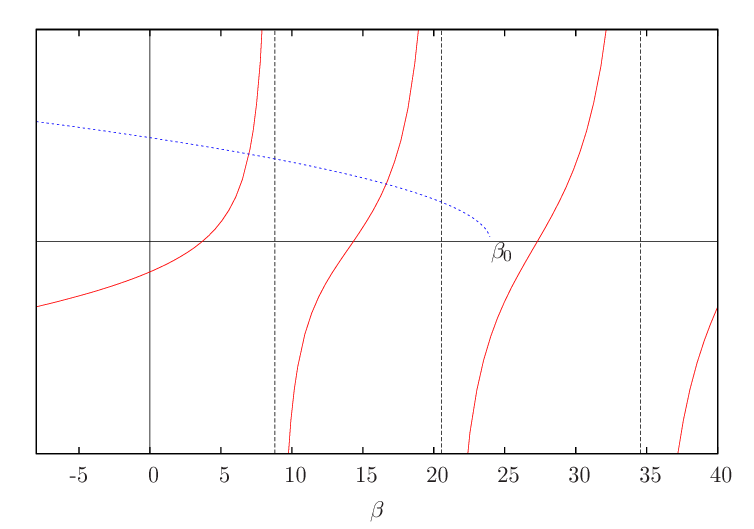}
\caption{Graphical solutions of \protect\eqref{energylevelond_exp}. The red solid line is the graph of $K_{i\sqrt{\beta}}'(\alpha)/K_{i\sqrt{\beta}}(\alpha)$, whereas the blue dash-dotted line represents $\sqrt{\beta_0 - \beta}/\alpha$ for the choice $\beta_0 = 24$ and $\alpha = 1$.}\label{Fig1_exp}
\end{center}
\end{figure}

If $E < U_0$ we obtain
\begin{equation}\label{WlU0sol_exp}
u(x) = \begin{cases}
CK_{i\sqrt{\beta}}\left(\alpha e^{-x/2\sigma}\right) & x \leq 0,\\
F e^{-k x} &  x > 0.
\end{cases}
\end{equation}
The junction conditions in $x = 0$ give
\begin{equation}\label{Condenlevel_exp}
\alpha K_{i\sqrt{\beta}}'(\alpha) = 2\sigma k K_{i\sqrt{\beta}}(\alpha) ,
\end{equation}
which, setting $\beta_0 := 8m(U_0 + \kappa)\sigma^2/\hbar^2$ can be recast in the form:
\begin{equation}\label{energylevelond_exp}
\frac{K_{i\sqrt{\beta}}'(\alpha)}{K_{i\sqrt{\beta}}(\alpha)} = \dfrac{\sqrt{\beta_0 - \beta}}{\alpha} .
\end{equation}
Graphical solutions of  \eqref{energylevelond_exp} are shown in \figurename~\ref{Fig1_exp}.
Analogously to what happens in the step-linear case \eqref{energylevelcond}, in the limit of an infinite barrier ($U_0 \to \infty$) the energy levels are specified by the zeros of $K_{i\sqrt{\beta}}(\alpha)$, the denominator of  \eqref{Condenlevel_exp}, as a function of $\beta$ and they are the ones of the symmetric confining potential $U(x) = \kappa\left(e^{|x|/\sigma}-1\right)$, corresponding to the odd eigenfunctions of the latter (Appendix~\ref{App:exponwell}).

To study how the energy levels behave for large energies, we employ the following formula for the asymptotic behavior of the function $K_{i\sqrt{\beta}}(\alpha)$ for large $\beta$ (see \cite{AS1972}):
\begin{equation}\label{asympexpr_exp}
K_{i\sqrt{\beta}}(\alpha) \sim \sqrt{\dfrac{2\pi}{\sqrt{\beta}}}e^{-\pi\sqrt{\beta}/2}\sin\left[\dfrac{\alpha^2}{4\sqrt{\beta}} - \sqrt{\beta} + \sqrt{\beta}\log\left(\dfrac{2\sqrt{\beta}}{\alpha}\right) + \dfrac{\pi}{4}\right]\left[1 + O\left(\dfrac{1}{\sqrt{\beta}}\right)\right] ,
\end{equation}
for $\beta \to \infty$ . Note that expansion \eqref{asympexpr_exp} can be proved starting from  \eqref{uzeq}. Therefore, the zeros of $K_{i\sqrt{\beta}}(\alpha)$, as a function of $\beta$, are asymptotically the solutions of the following equation:
\begin{equation}\label{asymptzeroes}
-\sqrt{\beta_n} + \sqrt{\beta_n}\log\left(\dfrac{2\sqrt{\beta_n}}{\alpha}\right) = n\pi .
\end{equation}
Solving for $\beta_n$ we obtain
\begin{equation}
 \beta_n \sim \frac{\alpha^2e^2}{4}\exp\left[2W\left(\frac{2n\pi}{\alpha e}\right)\right] ,
\end{equation}
where $W(x)$ is the Lambert function \cite{AS1972}. Since $W(x) \sim \log x - \log\log x$ for $x \to \infty$ we have for large $n$ that
\begin{equation}\label{expenlevelsasymp}
 \beta_n \sim \frac{n^2\pi^2}{\left(\log\frac{2n\pi}{\alpha e}\right)^2} .
\end{equation}
We see from  \eqref{expenlevelsasymp} that the potential $U(x) = \kappa\left(e^{|x|/\sigma} -1\right)$ behaves for large $x$ as an infinite square well whose width, up to inessential factors, grows as $\log n$, an intuitive fact. Moreover,
\begin{equation}\label{levspacingexp}
\beta_{n + 1} - \beta_{n} \sim \frac{2n\pi^2}{\left(\log\frac{2n\pi}{\alpha e}\right)^2} ,
\end{equation}
for $n\to\infty$, proving thus that the level spacing diverges.

\subsection{The case $E > U_0$: scattering and delay}

The unbound eigenstates have the form
\begin{equation}\label{WgU0sol_exp}
u(x) = \begin{cases}
CK_{i\sqrt{\beta}}(\alpha e^{-x/2\sigma}) & x \leq 0,\\
A e^{i kx} + B e^{-i kx} & x > 0.
\end{cases}
\end{equation}
and the junction conditions are
\begin{equation}\label{juncondsWgU0_exp}
\begin{cases}
CK_{i\sqrt{\beta}}(\alpha) = A + B,\\
C\alpha K'_{i\sqrt{\beta}}(\alpha) = i 2\sigma k\left(B - A\right).
\end{cases}
\end{equation}
Therefore, the normalized (with respect to $k$) improper eigenfunctions are given by:
\begin{equation}\label{improper_eigenfunctions_exp}
u_k(x) = \frac{1}{\sqrt{2\pi}} \begin{cases} \Pi[\beta(k)]K_{i\sqrt{\beta(k)}}\left(\alpha e^{-x/2\sigma}\right) & x\leq 0, \\
e^{ikx+i\delta(k)}+e^{-ikx} & x > 0,
\end{cases}
\end{equation}
where $\hbar k = \sqrt{2m(E - U_0)}$ and
\begin{equation}
\Pi(\beta) = 2\left[K_{i\sqrt{\beta}}(\alpha) + \dfrac{\alpha}{2ik\sigma}K'_{i\sqrt{\beta}}(\alpha) \right]^{-1} , \\ \qquad e^{i\delta(k)} = \dfrac{2ik\sigma K_{i\sqrt{\beta}}(\alpha) - \alpha K'_{i\sqrt{\beta}}(\alpha)}{2ik\sigma K_{i\sqrt{\beta}}(\alpha) + \alpha K'_{i\sqrt{\beta}}(\alpha)} .
\end{equation}
Hence,
\begin{equation}\label{phaseshift_exp}
\delta(k) = 2\arctan\left[\dfrac{\alpha}{2\sigma k}\dfrac{K_{i\sqrt{\beta(k)}}'(\alpha)}{K_{i\sqrt{\beta(k)}}(\alpha)}\right] .
\end{equation}
Then, following the same argument adopted for the step-linear case, we obtain the following formula for the delay $\tau$ of the rebound of an incoming wavepacket with peak energy $\beta$:
\begin{equation}\label{phaseshift_expbeta2}
\tau(\beta) = \frac{8m\sigma^2}{\hbar} \dfrac{d\delta}{d\beta} = \frac{8m\sigma^2}{\hbar}\dfrac{2}{\frac{\sqrt{\beta - \beta_0}}{\alpha} \frac{K_{i\sqrt{\beta}}(\alpha)}{K^\prime_{i\sqrt{\beta}}(\alpha)} + \frac{\alpha}{\sqrt{\beta - \beta_0}} \frac{K^\prime_{i\sqrt{\beta}}(\alpha)}{K_{i\sqrt{\beta}}(\alpha)}}\left[-\dfrac{1}{2\left(\beta - \beta_0\right)} + \dfrac{d}{d\beta}\log\dfrac{K'_{i\sqrt{\beta}}(\alpha)}{K_{i\sqrt{\beta}}(\alpha)}\right] .
\end{equation}
Using  \eqref{asympexpr_exp}, we obtain for large values of $\beta$
\begin{equation}\label{asymptratio}
 \dfrac{K'_{i\sqrt{\beta}}(\alpha)}{K_{i\sqrt{\beta}}(\alpha)} \sim -\frac{\sqrt{\beta}}{\alpha}\cot\left(-\sqrt{\beta} + \sqrt{\beta}\log\frac{2\sqrt{\beta}}{\alpha} + \frac{\pi}{4} \right) , 
\end{equation}
from which
\begin{equation}\label{hotterm}
 \dfrac{d}{d\beta}\log\dfrac{K'_{i\sqrt{\beta}}(\alpha)}{K_{i\sqrt{\beta}}(\alpha)} \sim \frac{1}{2\beta} - \frac{1}{\sin \left[2\left(-\sqrt{\beta} + \sqrt{\beta}\log\frac{2\sqrt{\beta}}{\alpha}+\frac{\pi}{4}\right)\right]}\frac{1}{\sqrt{\beta}}\log\frac{2\sqrt{\beta}}{\alpha} . 
\end{equation}
A comment is here in order. In general, taking the derivative of an asymptotic expansion with respect to the variable or a parameter may lead to wrong results. However, in our case this procedure can be justified using the integral representation of  \eqref{knu} (we leave this as an exercise for the interested reader).

Thus, plugging \eqref{hotterm} into \eqref{phaseshift_expbeta2} we obtain the asymptotic behavior of the delay time for large $\beta$'s, namely
\begin{equation}
 \tau(\beta) \sim \frac{8m\sigma^2}{\hslash}\frac{1}{\sqrt{\beta}}\log\left(\frac{2\sqrt{\beta}}{\alpha}\right) ,
\end{equation}
or, in terms of the energy of the particle
\begin{equation}\label{delaycalcasymptex}
 \tau(E) \sim \frac{2\sigma\log(2E/\kappa)}{\sqrt{2E/m}} .
\end{equation}
As expected, \eqref{delaycalcasymptex} coincides with the large energy value of the half period of the classical particle subjected to the confining potential $U(x) = \kappa\left(e^{|x|/\sigma} - 1\right)$.
\begin{figure}[htbp]
\centering
\includegraphics[width=0.33\columnwidth]{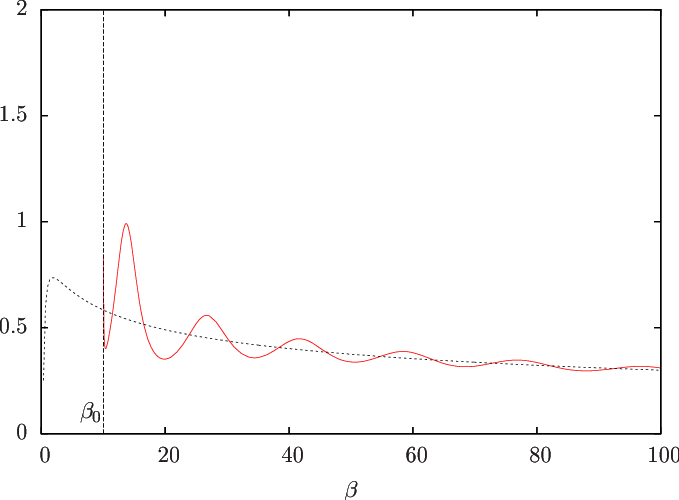}\includegraphics[width=0.33\columnwidth]{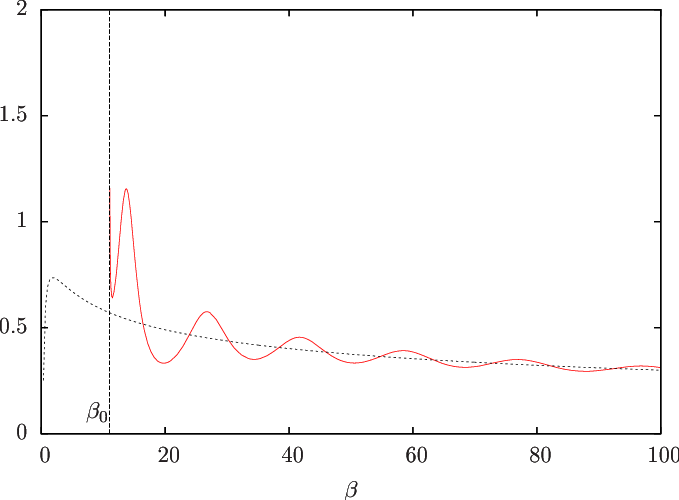}\includegraphics[width=0.33\columnwidth]{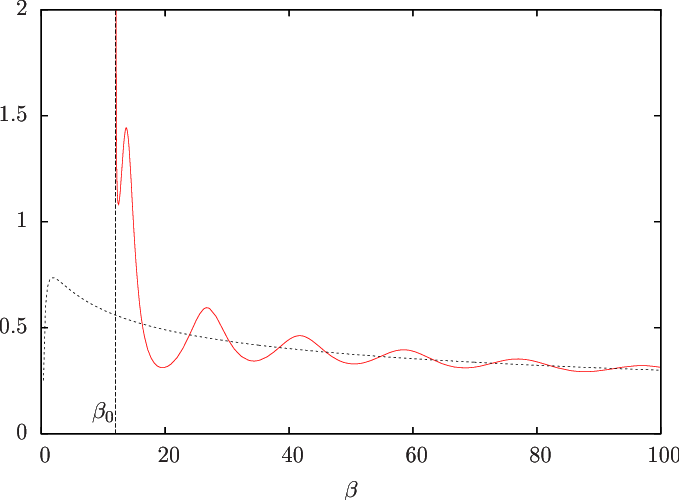}\\
\includegraphics[width=0.33\columnwidth]{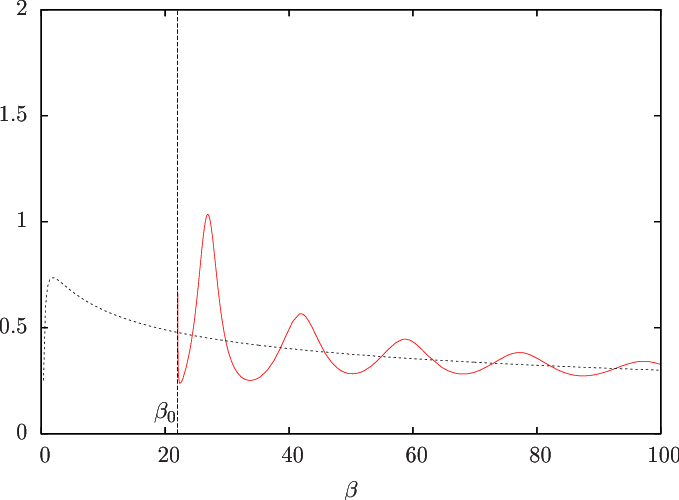}
\includegraphics[width=0.33\columnwidth]{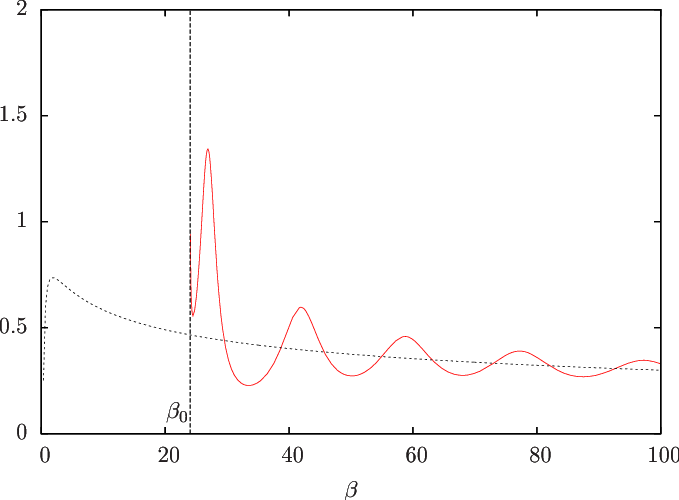}\includegraphics[width=0.33\columnwidth]{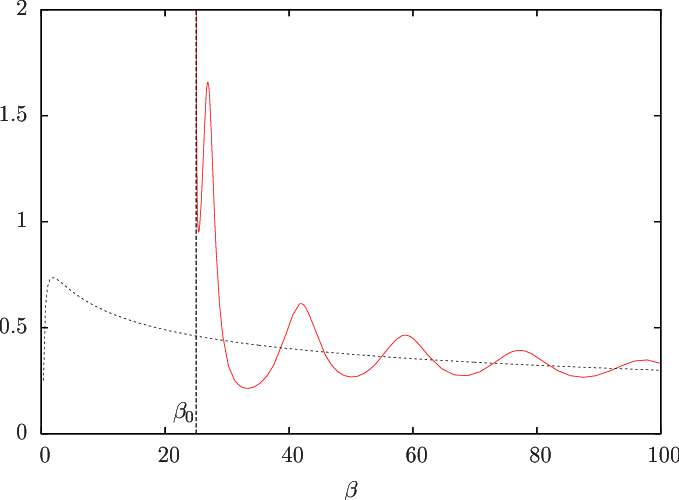}
\caption{Plots of the delay $\tau$ (in units of $8m\sigma^2/\hslash$) versus the energy $\beta$ of the incoming wave packet (solid red lines) for six values of the step height: $\beta_0 = 10, 11, 12$ (upper panels) $\beta_0 = 22, 24, 25$ (lower panels). The blue dotted lines represent the classical delay.}\label{Figrisexp}
\end{figure}

\section{Conclusions}\label{sec:Conclusions}

Regarding the structure of the discrete energy spectrum as a function of the height $U_0$ of the barrier, in the two potentials treated in this paper, the same considerations apply as those of the concluding section of \cite{RPCG2010}. The only difference is that the energy levels $\hslash\omega(n + 1/2)$, $n \in\mathbb{N}$, of the harmonic oscillator have to be replaced here by the corresponding levels $E_n$ of the confining linear and exponential potentials, respectively (see Appendices \ref{App:linearwell} and \ref{App:exponwell}). In the case of the step-linear potential, the level spacing \eqref{levspacinglin} goes to zero as the energy increases, while in the case of the step-exponential one \eqref{levspacingexp} it approaches infinity. As regards the continuous spectrum, we provide in both cases exact expressions for the delay of a wavepacket reflected from the barrier, as a function of the peak packet energy (see \eqref{phaseshiftbeta3} and \eqref{phaseshift_expbeta2}). As expected, in both cases these delays exhibit a series of resonances for energies not much larger than $U_0$, while for large energies, they approach the classical values. The step-harmonic potential is a threshold separating the potential barriers for which the delay time goes to infinity at large energies from those for which it vanishes. 

An entirely similar discussion can be applied to the step variant
\begin{equation}\label{stepvargeneral}
 U(x) = \begin{cases}
 V(x) & x \leq 0,\\
 U_0 & x > 0,
\end{cases}
\end{equation} 
of any symmetric potential $V(x)$ ($V(x) = V(-x)$) such that $\lim_{x\to\pm\infty}V(x)=+\infty$. Indeed the energy eigenvalue equation for $V(x)$ has two linearly independent solutions $u_L(x)$ and $v_L(x)$ the first of which approaches zero very rapidly as $x\to-\infty$ whereas the second one diverges steadily without oscillating, and two linearly independent solutions $u_R(x)$ and $v_R(x)$ having a corresponding behavior for $x\to+\infty$ (see \cite{Prosperi, Tricomi}). Since $u_L(x) = a(E)u_R(x) + b(E)v_R(x)$, the energy eigenvalues are the roots $E_n$ of the equation $b(E) = 0$. Since the potential is symmetric, these roots correspond to even and odd eigenfunctions alternatively, the ground state being even. However, in the general case the eigenvectors cannot be found explicitly. Therefore, for example, no explicit formula is available in general for the delay time of the reflected packet in the corresponding step variant potential \eqref{stepvargeneral}.

\section{Airy and modified Bessel functions through the integral representations}

In this section we solve the energy eigenvalue equations by means of the integral representation method, classifying the independent solutions as equivalence classes of homotopic paths in the complex plane. For the step-linear case we obtain Airy function, while for the step-exponential case we get modified Bessel functions. This technique is interesting \emph{per se}, as it can be applied to more general cases, provided one is able to guess the correct integral kernel. The Airy case is somehow classical, while the Bessel case is more interesting. We present them both for completeness.

\subsection{The step-linear case: Airy functions}\label{App:Airyfun}

We look for a solution of \eqref{Schreq3} of the form
\begin{equation}\label{prototipo2}
E(y) = \int_\gamma d t f(t)e^{ty} ,
\end{equation}
where $\gamma$ is a path in the complex plane and $f$ is an holomorphic function. Plugging \eqref{prototipo2} into \eqref{Schreq3} we find
\begin{equation}
\int_\gamma d t \left[(t^2 + \beta)f(t)e^{ty} + \frac{d e^{ty}}{d y} f(t)\right] = 0 .
\end{equation}
Integrating by parts, we obtain:
\begin{equation}
\left[e^{ty}f(t)\right]_{\partial\gamma} + \int_\gamma d t \left[(t^2 + \beta)f(t) - f(t)^\prime\right]e^{ty} = 0 .
\end{equation}
Therefore, $E(y)$ is a solution of \eqref{Schreq3} if
\begin{equation}
\left[e^{ty}f(t)\right]_{\partial\gamma} = 0 \qquad\text{and}\qquad f(t) = \exp\left(\frac{t^3}{3} + \beta t\right) .
\end{equation}
Hence, a class of solutions of the Airy equation is of the form
\begin{equation}\label{solclass}
E_\beta(y) = \int_\gamma d t \exp\left[\frac{t^3}{3} + (\beta + y)t\right] ,
\end{equation}
where $\gamma$ is a suitable path for which the contour term vanishes.

The integrand of \eqref{solclass} entire. Thus, by Cauchy theorem, every closed path represents the trivial solution $E(y) = 0$.

\begin{figure}
 \centering
 \includegraphics[width=0.45\columnwidth]{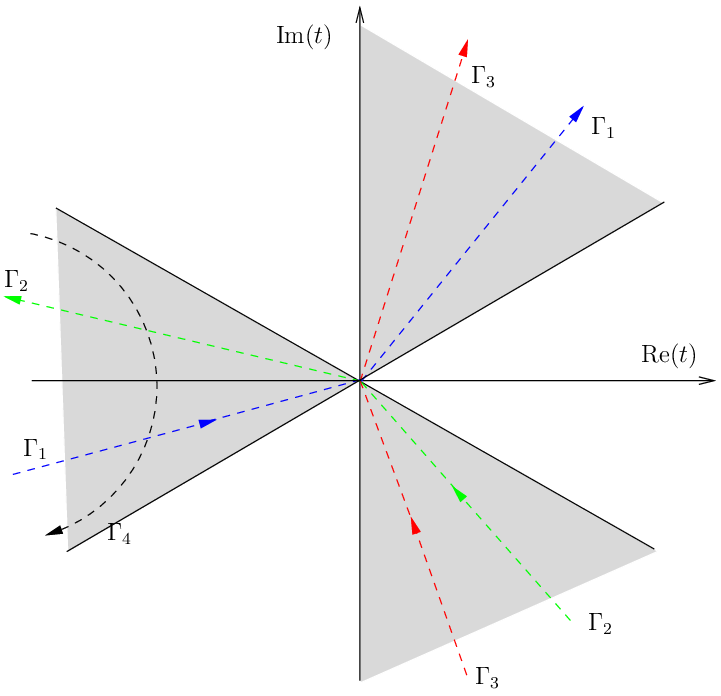}
 \caption{Possible paths of integration, each one corresponding to a solution of \protect\eqref{Schreq3}.} \label{fig:shaded}
\end{figure}

Consider an unbounded path. In order for $\left[e^{ty}f(t)\right]_{\partial\gamma}$ to vanish, we require the leading term in the exponent of $f(t)$ (i.e. $t^3$) to have a negative real part. Therefore, the acceptable unbounded paths are those whose phase $\phi$ is confined to the regions $\frac{\pi}{6} < \phi + \frac{2}{3}n\pi < \frac{\pi}{2}$ ($n = 0, 1, 2$). These possible paths are showed in \figurename~\ref{fig:shaded}, where the allowed sectors $\frac{\pi}{6} < \phi + \frac{2}{3}n\pi < \frac{\pi}{2}$ ($n = 0, 1, 2$) are shaded.

Paths with both endpoints in the same sector (e.g. $\Gamma_4$ in \figurename~\ref{fig:shaded}) can be closed at infinity using Jordan's Lemma; therefore, they correspond to the trivial solution. The only non-trivial paths are those which link different sectors. There are only 3 non-equivalent classes of such paths which we dub $\Gamma_1$, $\Gamma_2$ and $\Gamma_3$ respectively (see \figurename~\ref{fig:shaded}).
Taking into account Cauchy theorem, these paths satisfy the relation $\Gamma_1 + \Gamma_2 = \Gamma_3$ in the sense that the corresponding solutions are not independent. The conventional Airy functions $\Ai(z)$ and $\Bi(z)$ are the independent solutions of $w^{\prime\prime}(z) - z w(z) = 0$ such that (see \cite{AS1972})
\begin{equation}
\mathrm{Ai}(0) = \dfrac{3^{-2/3}}{\Gamma(2/3)}, \quad \mathrm{Ai}^\prime(0) = -\frac{3^{-1/3}}{\Gamma(1/3)}, \qquad \text{and}\qquad
\mathrm{Bi}(0) = \sqrt{3}\frac{3^{-2/3}}{\Gamma(2/3)}, \quad \mathrm{Bi}^\prime(0) = \sqrt{3}\frac{3^{-1/3}}{\Gamma(1/3)} .
\end{equation}
Denoting by $E^{(i)}_\beta(y)$ the solutions in \eqref{solclass} corresponding to the paths $\Gamma_i$ ($i = 1, 2$), it is not difficult to show that
\begin{equation}
\Ai(-y-\beta) = \frac{1}{2\pi i}\left[E^{(1)}_\beta(y) + E^{(2)}_\beta(y)\right] ,\qquad \Bi(-y-\beta) = \frac{1}{2\pi}\left[E^{(1)}_\beta(y) - E^{(2)}_\beta(y)\right] .
\end{equation}
We leave the details to the interested reader (\emph{hint}: Check the above expressions and their first derivatives in \eqref{solclass} for $y = \beta = 0$. In this case the integrals $E_0^{(j)}(0)$ correspond to Euler Gamma functions).

\subsection{The step-exponential case: modified Bessel functions}\label{App:modbesselfun}

\begin{figure}
\centering
\includegraphics[width=0.45\columnwidth]{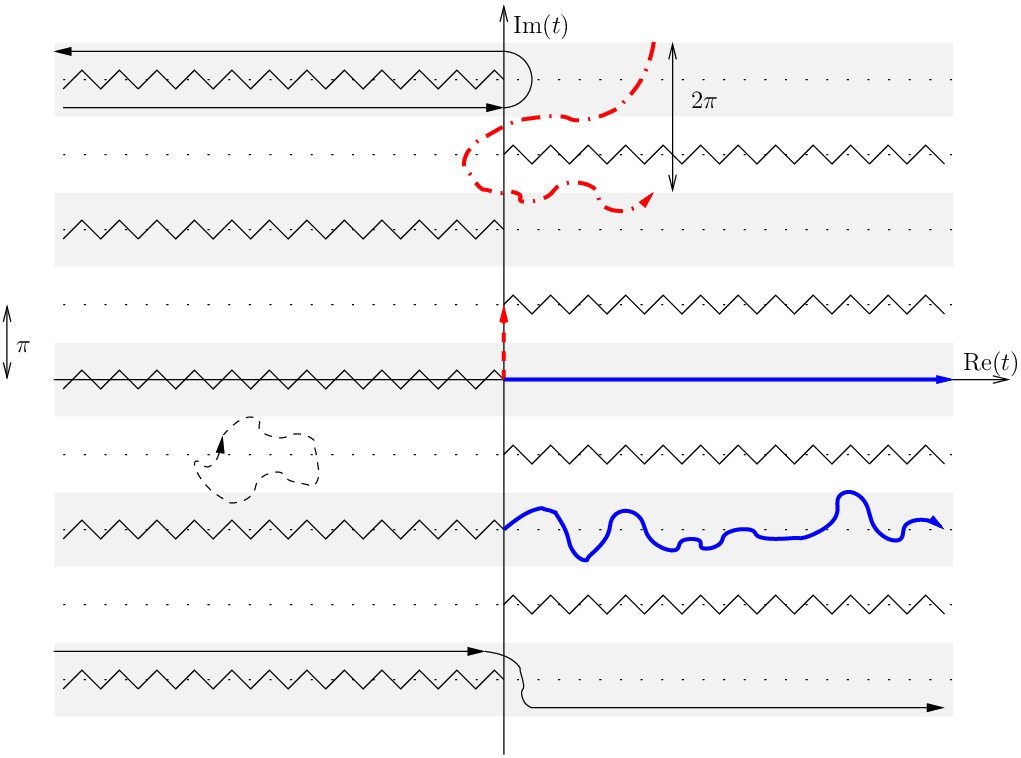}
\caption{Cut of the complex $t$-plane. The dashed red and the solid blue thick paths represent the only classes of paths which provide independent solutions to the modified Bessel equation.}\label{fig:modbessel_cuts}
\end{figure}

Consider the modified Bessel equation
\begin{equation}\label{eqn:modifiedbessel}
z^2 u^{\prime\prime}(z)+zu^\prime(z) -(\nu^2+z^2)u(z) = 0 ,
\end{equation}
with $z > 0$. A convenient kernel for the integral representation is the following:
\begin{equation}
K(z,t) = z^\nu e^{-z\cosh t} .
\end{equation}
We look for solutions of the form
\begin{equation}\label{eqn:candidate_modbessel}
u(z) = \int_\gamma d t  f(t)K(z,t) .
\end{equation}
Plugging \eqref{eqn:candidate_modbessel} into \eqref{eqn:modifiedbessel}, we get
\begin{equation}\label{intcondexp}
z^{\nu + 1}\int_\gamma d t \left[f(t)\frac{d}{d t}\left(\sinh(t)e^{-z\cosh t}\right) + 2\nu f(t) \cosh(t)e^{-z\cosh t}\right] = 0 .
\end{equation}
Integrating by parts, \eqref{intcondexp} gives
\begin{equation}
\left[z^{\nu + 1}e^{-z\cosh t}f(t)\sinh(t)\right]_{\partial\gamma} - z^{\nu + 1}\int_\gamma d t \left[f^{\prime}(t)\sinh(t) -2\nu f(t)\cosh(t)\right]e^{-z\cosh t} = 0 .
\end{equation}
Then, the integral on the right hand side of \eqref{eqn:candidate_modbessel} is a solution of \eqref{eqn:modifiedbessel} if
\begin{equation}
f^{\prime}(t)\sinh(t) -2\nu f(t)\cosh(t) = 0 , \qquad\mbox{ and }\qquad z^{\nu+1} \left[e^{-z\cosh t}f(t) \sinh(t)\right]_{\partial\gamma} = 0 .
\end{equation}
Up to a normalization, the solution is $f(t) = \sinh(t)^{2\nu}$. If $\nu$ is integer then $f(t)$ is either entire ($\nu \geq 0$) or meromorphic ($\nu < 0$), otherwise it has infinite branch points located at $t_n = i n \pi$ ($n = 0, \pm1, \dots$), see \figurename~\ref{fig:modbessel_cuts}. In the latter instance, the usual procedure is to define a domain in which the function is holomorphic by cutting the $t$-plane and thus forbidding loops around the branch points. In \figurename~\ref{fig:modbessel_cuts} a convenient choice for the cuts is also shown.
Recalling that $\mathrm{Re}(z) > 0$, the contour condition $z^{\nu+1} \left[e^{-z\cosh(t)}f(t) \sinh(t)\right]_{\partial\gamma} = 0$ is
\begin{equation}\label{contcond}
 e^{-z\cosh t}\sinh(t)^{2\nu + 1} \rvert_{\partial\gamma}= 0 ,
\end{equation}
There are 4 different classes of paths for which \eqref{contcond} is satisfied and
\begin{equation}\label{uzeq}
u(z) = \int_\gamma \sinh(t)^{2\nu} e^{-z\cosh t}z^\nu dt ,
\end{equation}
is well defined. The ``paths zoology'' is more complicated and rich than in the linear and in the harmonic case \cite{RPCG2010}.
\begin{enumerate}
\item {\bf Closed paths.} For any closed path, the contour condition is trivially satisfied and any integral along a path enclosing a region where the integrand function is holomorphic (i.e. the path does not cross the cuts) vanishes. An example is shown by the thin dashed black line in \figurename~\ref{fig:modbessel_cuts}.

\item {\bf Infinite paths.} For paths whose endpoints are both at infinity, the function $\sinh(t)^{2\nu + 1}$ diverges or oscillates. On the other hand, the exponential $e^{-z\cosh t}$ vanishes for $\mathrm{Re}(\cosh t)\rightarrow +\infty$ (recall that $z > 0$). Since $\mathrm{Re}[\cosh(x+iy)] = \cos(y)\cosh(x)$ then $\gamma$ must stretch at infinity in one of the sectors defined by $-\pi/2 + 2n\pi < \mathrm{Im}(t) < \pi/2 + 2n\pi$ ($n = 0, \pm 1, ...$), which are represented by the shaded regions in \figurename~\ref{fig:modbessel_cuts}. Incidentally, in these bands, when there are no cuts, one can ``close'' the paths at infinity, by virtue of Jordan's Lemma. Examples of this class of paths are the black solid thin lines of \figurename~\ref{fig:modbessel_cuts}.

\item {\bf Semi-infinite paths.} By ``semi-infinite'' paths we mean paths starting from a point, say $t_0$, and ending at infinity. These paths must go to infinity in the shaded bands $-\pi/2 + 2n\pi < \mathrm{Im}(t)< \pi/2 + 2n\pi$. For the starting point $t_0$, the contour condition demands that $\sinh(t_0)^{2\nu + 1} = 0$. This means that these paths must start from one of the points $t_n = i n\pi$, which are the zeroes of the hyperbolic sine function. It is easy to prove that the integral of \eqref{uzeq} performed along any two such paths lying in the same band gives the same result (indeed, recall that $\sinh t$ is periodic). Two examples of this class of paths are the blue solid thick lines in \figurename~\ref{fig:modbessel_cuts}.

\item {\bf Finite paths.} These are the paths starting and ending in two points say $t_i$ and $t_f$, with $\abs{t_i},\abs{t_f} < \infty$. The contour condition can be satisfied in two different ways: either the values of the contour part are equal at the endpoints, or the contour part vanishes at the endpoints. The former case accounts for paths which do not cross any cut and start from any point $t_i$, ending at $t_f = t_i + 2 n i\pi$ ($n = 0, \pm1, \dots$). Examples of this class of paths are represented by the red dash-dotted thick line in
\figurename~\ref{fig:modbessel_cuts}.
The latter case is realized by paths connecting the branch points and is represented by the red dashed thick line in \figurename~\ref{fig:modbessel_cuts}.
\end{enumerate}
Taking into account the periodicity of the integrand function (the period is $2\pi i$ in the domain where it is holomorphic), and the Cauchy theorem, it is easy to show that the integrals along the two kinds of finite paths are proportional to the integral along the ``fundamental'' path $[0,i\pi]$. Moreover, the integrals along any one of the infinite paths are linear combinations of the ones performed along the finite and semi-infinite paths.

In conclusion, two linear independent solutions of \eqref{eqn:modifiedbessel} are
\begin{equation}\label{knu}
K_\nu(z) = \frac{\pi^{1/2}(z/2)^\nu}{\Gamma(\nu + 1/2)}\int_0^{\infty}d x e^{-z\cosh x}\sinh(x)^{2\nu} ,
\end{equation}
and
\begin{equation}\label{inu}
I_\nu(z) = \frac{(z/2)^\nu}{\pi^{1/2}\Gamma(\nu + 1/2)}\int_0^{\pi}d x e^{-z\cos x}\sin(x)^{2\nu} ,
\end{equation}
where $K_\nu$ and $I_\nu$ correspond, respectively, to the integrals performed along the solid blue and the dashed red thick lines in Fig.~\ref{fig:modbessel_cuts} (a semi-infinite path and a finite one). It can be shown that both solutions (derived here for $z > 0$) can be analytically continued throughout the whole $z$-plane cut along the negative real axis (see \cite{Hochstadt1976}).

\appendix

\section{The confining symmetric linear potential}\label{App:linearwell}

Consider the confining symmetric potential  $U(x) = M|x|$. The eigenfunctions can be written as
\begin{equation}
\begin{cases}
u(x) = C_1 {\rm Ai}(-\alpha x - \beta) &  x < 0,\\
u(x) = C_2 {\rm Ai}(\alpha x - \beta) &  x > 0,
\end{cases}
\end{equation}
where $C_1$ and $C_2$ are constants fixed by the junction conditions in $x = 0$. If ${\rm Ai}(- \beta) \neq 0$, then the continuity of $u(x)$ in $x = 0$ implies $C_1 = C_2$. Moreover, the continuity of the derivative implies ${\rm Ai}^\prime(- \beta) = 0$. This condition determines the \emph{even} eigenfunctions and their eigenvalues.
If ${\rm Ai}(- \beta) = 0$, then the continuity of the derivative implies $C_1 = - C_2$. This condition determines the \emph{odd} eigenfunctions and their eigenvalues.

\section{The confining symmetric exponential potential}\label{App:exponwell}

Consider the confining symmetric potential $U(x) = \kappa\left(e^{|x|/\sigma} -1\right)$. The eigenfunctions can be written as
\begin{equation}
\begin{cases}
 u(x) = C_1 K_{i\sqrt{\beta}}(\alpha e^{-x/2\sigma}) &  x < 0,\\
u(x) = C_2 K_{i\sqrt{\beta}}(\alpha e^{x/2\sigma}) & x > 0,
\end{cases}
\end{equation}
where $C_1$ and $C_2$ are constants fixed by the junction conditions in $x = 0$. If $K_{i\sqrt{\beta}}(\alpha) \neq 0$, then the continuity of $u(x)$ in $x = 0$ implies $C_1 = C_2$. Moreover, the continuity of the derivative implies $K_{i\sqrt{\beta}}^\prime(\alpha) = 0$. This condition determines the \emph{even} eigenfunctions and their eigenvalues.
If $K_{i\sqrt{\beta}}(\alpha) = 0$, then the continuity of the derivative implies $C_1 = - C_2$. This condition determines the \emph{odd} eigenfunctions and their eigenvalues.

\end{document}